# ALPHA-SPECTROSCOPY OF $^{252}$Cf DECAYS: A NEW APPROACH TO SEARCHING FOR THE OCTONEUTRON

Hrant Gulkanyan and Amur Margaryan

A.I. Alikhanyan National Science Laboratory (Yerevan Physics Institute), Yerevan, Armenia

**Abstract**

Recently an experimental evidence of emission of neutron clusters consisting of eight neutrons - octoneutrons was reported. The experiment was based on the search for daughter nuclei associated with the octoneutron radioactivity of the $^{252}$Cf nucleus. The effect of long-time build-up of daughter nuclei with a long half-life and gamma spectroscopy were utilized. In this paper we consider the alpha spectroscopy as an alternative method of searching for octoneutron in $^{252}$Cf decays.

1. **Introduction**

The existence of stable nuclei composed of only neutrons is one of the important problems of strong interaction and nuclear physics. The discovery of such many-neutron systems will induce essential progress in our understanding of the evolution of the Universe and the cosmic nucleosynthesis. In the experiments of last decades, aimed at the search for dineutron ($^2$n), trineutron ($^3$n) and tetraneutron ($^4$n) in various nuclear reactions, no decisive evidence for the existence of those systems was obtained. However, the possibility of the existence of heavier neutron nuclei is not excluded [1, 2, 3]. Probably, more favorable conditions for their formation can be provided in the reactions of spontaneous and induced fission of heavy, neutron-rich nuclei.

Recently a new method to look for the emission of many-neutron systems (clusters) from heavy nuclei was proposed [3]. The method is based on the detection of decays of long-living radioactive nuclides associated with the cluster radioactivity of heavy (transuranium) nuclei. With this method, indirect evidence was obtained [3] for the octoneutron ($^8$n) radioactivity of the $^{252}$Cf nucleus. It should be, however, pointed out that the results of [3] can have alternative interpretations too, and new detailed experimental investigations are needed to get more direct and reliable evidences on the manifestation of the octoneutrons. In this Proposal, we suggest more direct method for identification of the octoneutron radioactivity of the $^{252}$Cf nucleus.

## 2. Physics motivation

The octoneutron emission from $^{252}$Cf is an endothermic reaction

$$^{252}\text{Cf} \rightarrow (^8\text{n}) + {}^{244}\text{Cf} \qquad (1)$$

which can proceed only if the octoneutron binding energy exceeds 50 MeV. This reaction is followed by a chain of more than ten α- and β-deacys (see Figure 1) which ends with a lead nucleus being produced either due to the α-decay of $^{212}$Po (with a probability of 64%) or due to the β-decay of $^{208}$Tl (with a probability of 36%). The latter is almost always (in 99.8% of decays) accompanied by the emission of a $E_\gamma$ = 2615 keV γ-quantum – a process which can last very long (more than one hundred years) even if the content of the mother isotope $^{252}$Cf in the radioactive source is practically fully exhausted due to the "conventional" α-decay and spontaneously fission processes. This happens owing to accumulation of the long-living uranium isotope $^{232}$U (with half-life of $T_{1/2}$ = 68.9 years) whose decay chain just provides a long-term surviving of this, delayed γ-radiation signature of the octoneutron radioactivity (see Figure 1). It should be pointed out here that an accumulation of $^{232}$U nuclei can also take place due to another (not yet observed experimentally) channel of the cluster radioactivity of $^{252}$Cf,

$$^{252}\text{Cf} \rightarrow {}^{20}\text{C} + {}^{232}\text{U}, \qquad (2)$$

the theoretically predicted yield of which is, however, too low to fit the measured [3] intensity of the $E_\gamma$ = 2615 keV line.

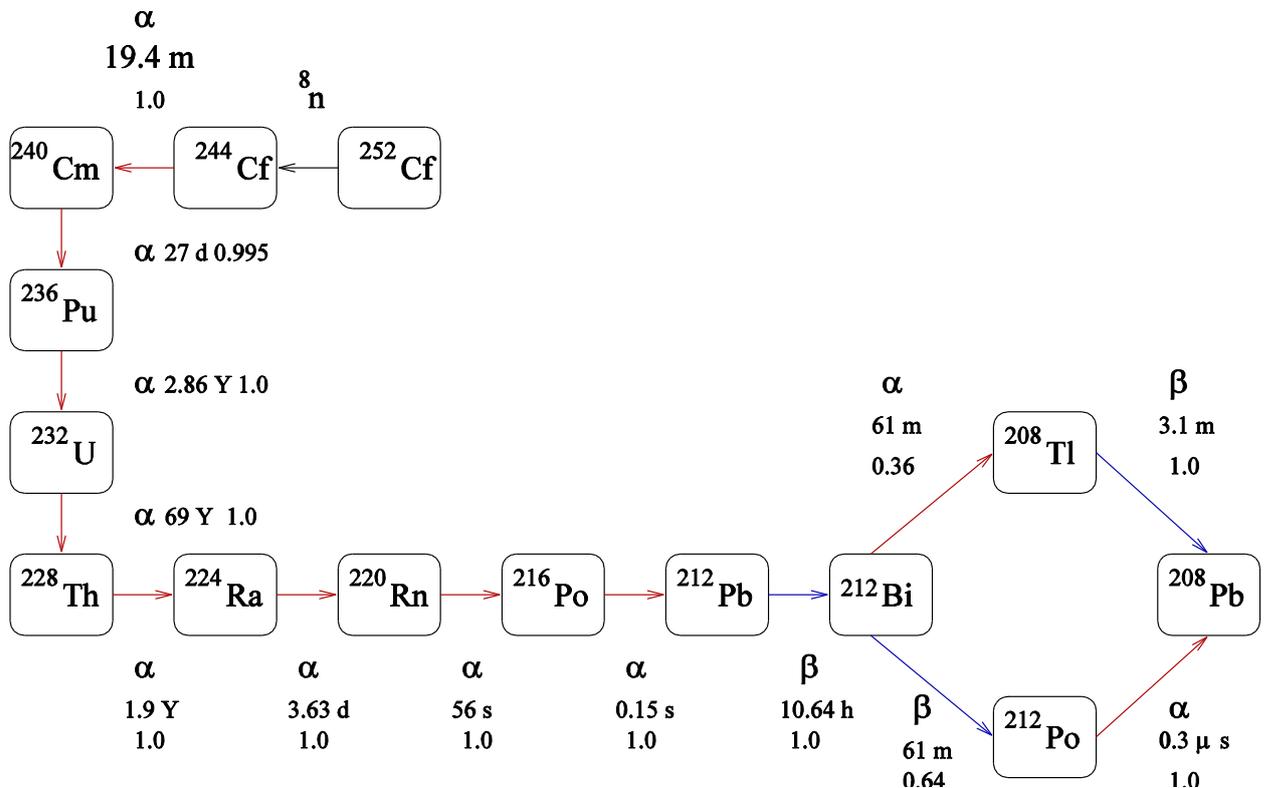

Fig. 1. The decay chain followed from the hypothetical octoneutron emission by $^{252}$Cf. α- and β-decays and corresponding half-lives are indicated (Y - year, d - day, h - hour, m - minute, s - second). The corresponding decay probabilities are also quoted (1.0, 0.995, 0.36, 0.64). The data are taken from www.nndc.bnl.gov.

According to estimations [3], the role of other possible processes in the emission of $E_\gamma = 2615$ keV photons is also negligible, and the measured intensity of this spectral line can be almost completely ascribed to the reaction (1), which yield relative to that of α-decay is inferred to be $P(^8n)/P(\alpha) = 1.74 \cdot 10^{-6}$. As it was pointed out in [3], in order to prove the existence of the octoneutron radioactivity, similar independent experiments should be performed using $^{252}$Cf sources of different ages (including fresh sources) which are characterized by different fractions of accumulated $^{232}$U nuclei.

In this Project, a new, more direct method is suggested for identification of the decay chain presented in Fig. 1. The method is based on the measurement of spectral characteristics of α-particles associated with the decay chain. Just at the first link of the chain, the daughter nucleus $^{244}$Cf emits α-particle with energy of $E_\alpha = 7.2$ MeV; such energy is not possessed either by α-particles from the subsequent links of the chain or by α-particles from the main decay chain initiated by the α-decay of $^{252}$Cf (see Fig. 2; the corresponding α-particle energies in both chains are listed in Table 1).

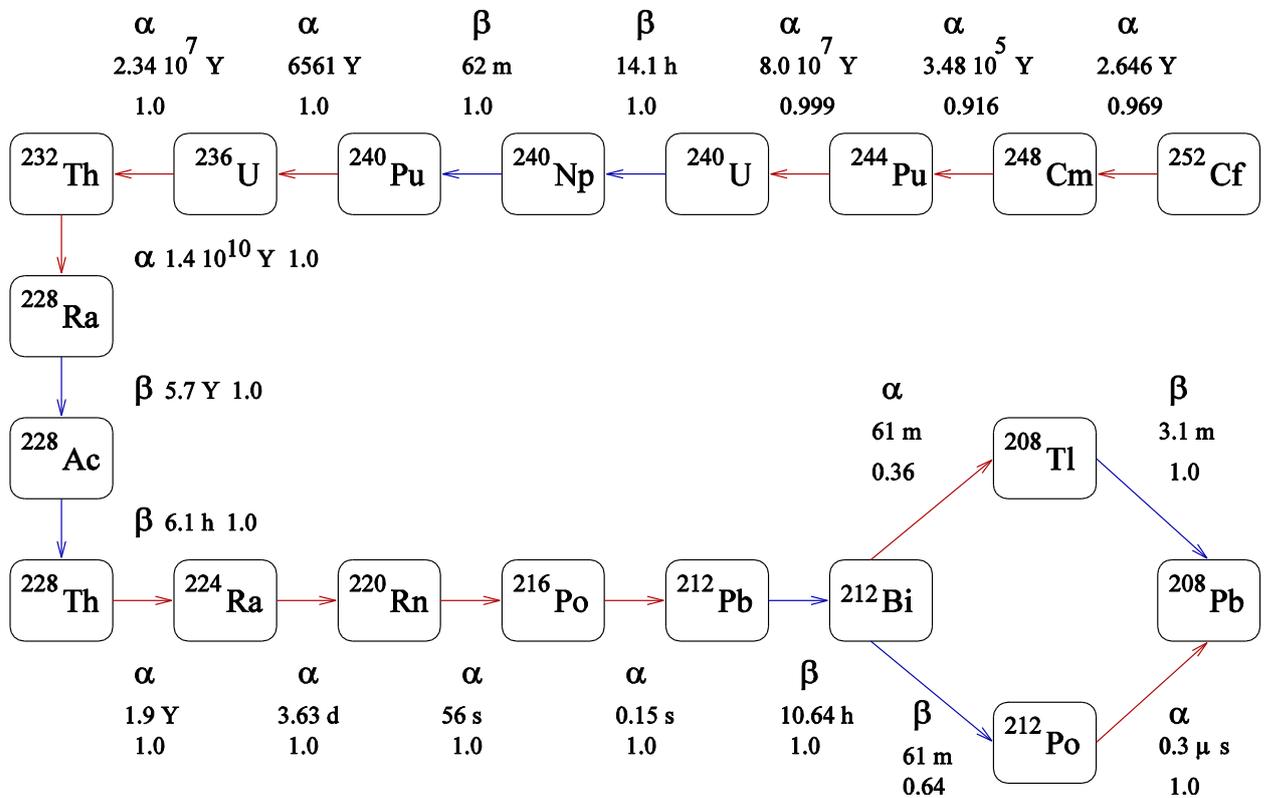

Fig. 2. The decay chain followed from the α-decay of $^{252}$Cf. The notation is the same as in Fig. 1.

It should be pointed out that, in the case of the existence of the process (1), the line $E_\alpha = 7.2$ MeV has to be certainly accompanied by other, relatively high-energy α-lines from the decays of $^{216}$Po ($E_\alpha = 6.78$ MeV) and $^{212}$Po ($E_\alpha = 8.78$ MeV), as it can be seen from Fig. 1 and Table 1. Moreover, for the case of sufficiently old $^{252}$Cf source, the intensity of these lines can exceed significantly that for the $E_\alpha = 7.2$ MeV line due to the aforementioned accumulation effect for $^{232}$U nuclei (see numerical values below in Section 4). Hence, a simultaneous observation of mentioned three high-energy α-lines (with characteristic relative proportions, see below) in the α-particle spectrum, can serve as a direct evidence for the chain depicted in Fig. 1, while the absence of at least one of these three lines in the α-spectrum from a radioactive source containing a sufficient amount of $^{252}$Cf will exclude (within measurement errors) the existence of the reaction (1). On the other hand, at the absence of the $E_\alpha = 7.2$ MeV line and, simultaneously, presence of other two lines in the α-spectrum will mean that in the interpretation of the experimental results [3] one should not disregard the role of reaction (2), as well as the following cluster radioactivity reactions (not yet observed experimentally), resulting in the production of both $^{216}$Po and $^{212}$Po daughter isotopes:

$$^{252}\text{Cf} \rightarrow {}^{24}\text{O} + {}^{228}\text{Th}, \qquad (3)$$

$$^{252}\text{Cf} \rightarrow {}^{28}\text{Ne} + {}^{224}\text{Ra}, \qquad (4)$$

$$^{252}\text{Cf} \rightarrow {}^{32}\text{Mg} + {}^{220}\text{Rn}, \qquad (5)$$

$$^{252}\text{Cf} \rightarrow {}^{36}\text{Si} + {}^{216}\text{Po} \qquad (6)$$

Table 1. The essential α-lines in decay chains presented in Figs. 1 and 2.

| Mother nucleus | Energy (keV) | Intensity (%) |
|---|---|---|
| $^{252}$Cf → $^{244}$Cf + $^{8}$n | | |
| $^{244}$Cf | 7209 | 53 |
| | 7174 | 18 |
| $^{240}$Cm | 6291 | 70.9 |
| | 6248 | 28.8 |
| $^{236}$Pu | 5768 | 69.1 |
| | 5721 | 30.8 |
| $^{232}$U | 5320 | 68.2 |
| | 5263 | 31.6 |
| $^{228}$Th | 5423 | 72.2 |
| | 5340 | 27.2 |
| $^{224}$Ra | 5685 | 94.9 |
| $^{220}$Rn | 6288 | 99.9 |
| $^{216}$Po | 6778 | 100 |
| $^{212}$Bi | 6051 | 25.1 |
| $^{212}$Po | 8785 | 100 |
| $^{252}$Cf → $^{248}$Cm + α | | |
| $^{252}$Cf | 6118 | 81.6 |
| | 6076 | 15.2 |
| $^{248}$Cm | 5078 | 75 |
| | 5035 | 16.5 |
| $^{244}$Pu | 4589 | 80.5 |
| | 4546 | 19.4 |
| $^{240}$Pu | 5168 | 72.8 |
| | 5124 | 27.1 |
| $^{236}$U | 4497 | 74 |
| | 4445 | 26 |
| $^{232}$Th | 4012 | 78.2 |
| | 3947 | 21.7 |

Further, in the case of the simultaneous absence of the first two lines and the presence the $^{212}$Po α-line ($E_α$ = 8.78 MeV), the following, not yet observed, cluster radioactivity reactions, bypassing the production of the $^{216}$Po isotope, have to be considered:

$$^{252}\text{Cf} \rightarrow {}^{40}\text{S} + {}^{212}\text{Pb} \tag{7}$$

$$^{252}\text{Cf} \rightarrow {}^{40}\text{P} + {}^{212}\text{Bi}, \tag{8}$$

while, at the absence of both three lines, the emission of the $E_γ$ = 2615 keV γ-quanta can be contributed from the following hypothetical cluster emission reactions:

$$^{252}\text{Cf} \rightarrow {}^{44}\text{Cl} + {}^{208}\text{Tl}, \tag{9}$$

$$^{252}\text{Cf} \rightarrow {}^{44}\text{P} + {}^{208}\text{Bi}. \tag{10}$$

One can, therefore, conclude that a detailed investigation of the energy spectrum of α-particles from a $^{252}$Cf radioactive source can be decisive for the proof of the octoneutron radioactivity.

## 3. Proposed experimental device

We propose to implement the α-particle spectroscopy of the $^{252}$Cf nucleus to confirm the existence of its octoneutron radioactivity. The main goal of the proposed experiment is to detect the relatively high-energy tail ($E_α$ > 6.5 MeV) of α-particles corresponding to the decays of $^{244}$Cf, $^{216}$Po and $^{212}$Po initiated from the hypothetical reaction (1). In principle, α-particle spectroscopy can be accomplished by using a Si detector to measure energies of α-particles, which can be realized with about 100 keV resolution or better. However, as the relative probability of the decay channel, which we are interested in, is very small, it is worthy to consider a more sophisticated device, which will be able to decrease the background from the $^{252}$Cf fission fragments (FFs) and cosmic rays.

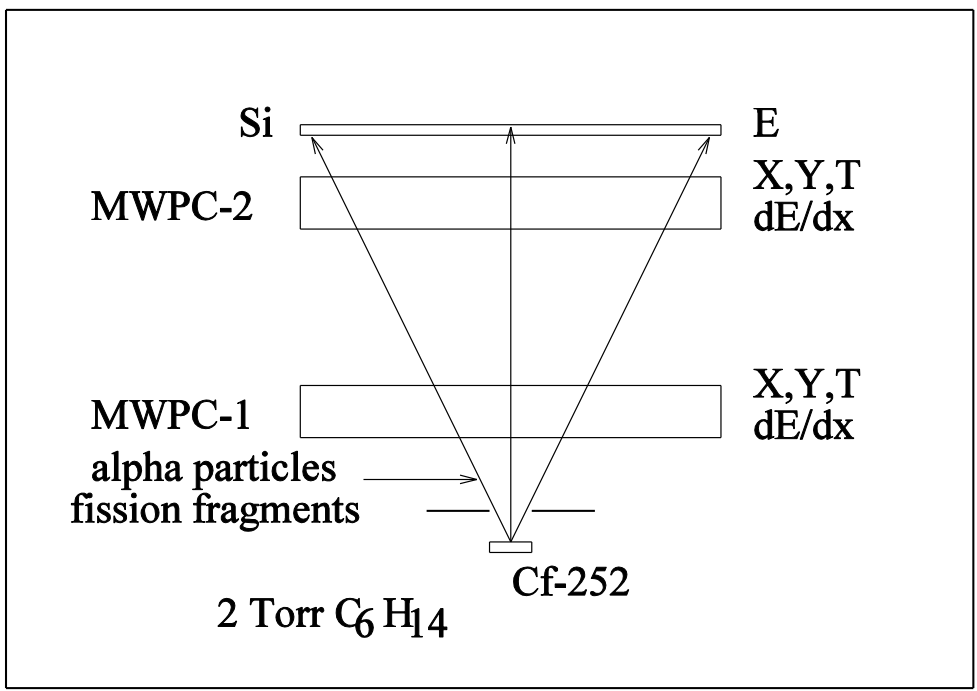

Fig. 3. A sketch of the proposed experimental setup.

To minimize the mentioned background, we propose an experimental setup which is schematically depicted in Fig. 3. It consists of two low-pressure MWPC units and a Si detector. Each MWPC unit will provide X and Y coordinates, t – time and dE/dx, the ionization energy losses of α- particles and heavier nuclear fragments from $^{252}$Cf. The Si detector will measure the kinetic energies of these particles. Using this information, the trajectories of decay particles, their velocities and masses will be reconstructed. By this way the background related to FFs and cosmic rays will be minimized. Thus, the proposed device will allow to carry out a sensitive search for $^{252}$Cf rare decays, were 6.6 ÷ 8.8 MeV α-particles are produced.

**4. Preliminary investigations**

We have developed and tested a low-energy α-particle detector based on the low-pressure MWPCs for nuclear fission and fragmentation studies planned at MAX-lab [4] and at the Yerevan Physics Institute.

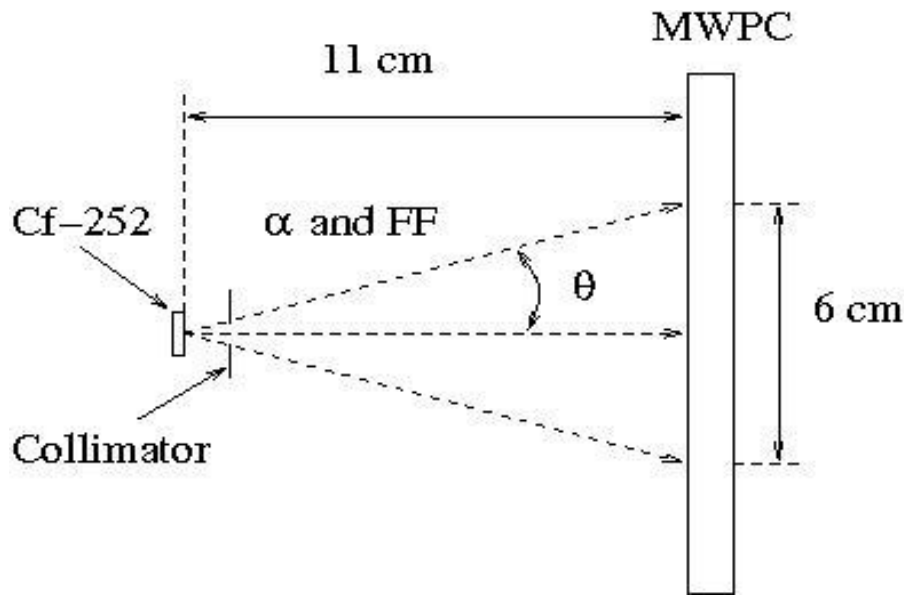

Fig. 4. A sketch of the MAX-lab test experiment. The figure is taken from [4].

In Fig. 4 the experimental setup, which was used at MAX-lab [4] to measure ionization energy losses of α-particles and FFs, is presented. The geometry of the experimental setup allows to measure dE/dx of particles with θ angles less than 15 degrees. The distribution of amplitudes of signals generated by α-particles is shown in Fig. 5. This distribution has the expected Landau distribution shape and can be described by using Monte Carlo calculations based on the individual collision method [5].

The signals generated by FFs are by from several tens to hundred (in average 80) times larger than signals generated by α-particles [4]. On the other hand, the low-pressure MWPCs are insensitive to cosmic rays, β and γ radiations.

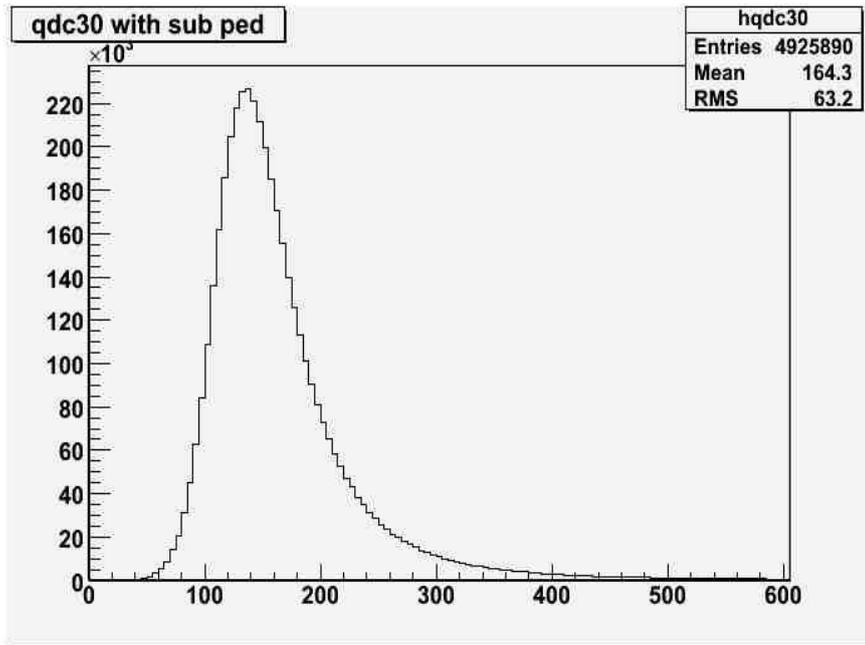

Fig. 5. The distribution of signals generated by α-particles from Cf-252 in the low-pressure MWPC. The horizontal scale is in arbitrary units. The figure is taken from [4].

The dE/dx (a) and energy (b) distributions of α-particles from $^{252}$Cf (6.118 MeV, 81.6% and 6.076 MeV, 15.2%) passing through 7.5 (75) µg/cm$^2$ carbon, which is equivalent to about 2 Torr and 1(10) cm hexane, simulated by means of individual collision method, are presented in Fig. 6 and Fig. 7. The geometry of these simulations is similar to the geometry depicted in Fig. 4. In addition, Fig. 7 presents the distributions of energies of α-particles detected with 30 keV(Fig. 7c) and 100 keV(Fig. 7d) resolutions, respectively.

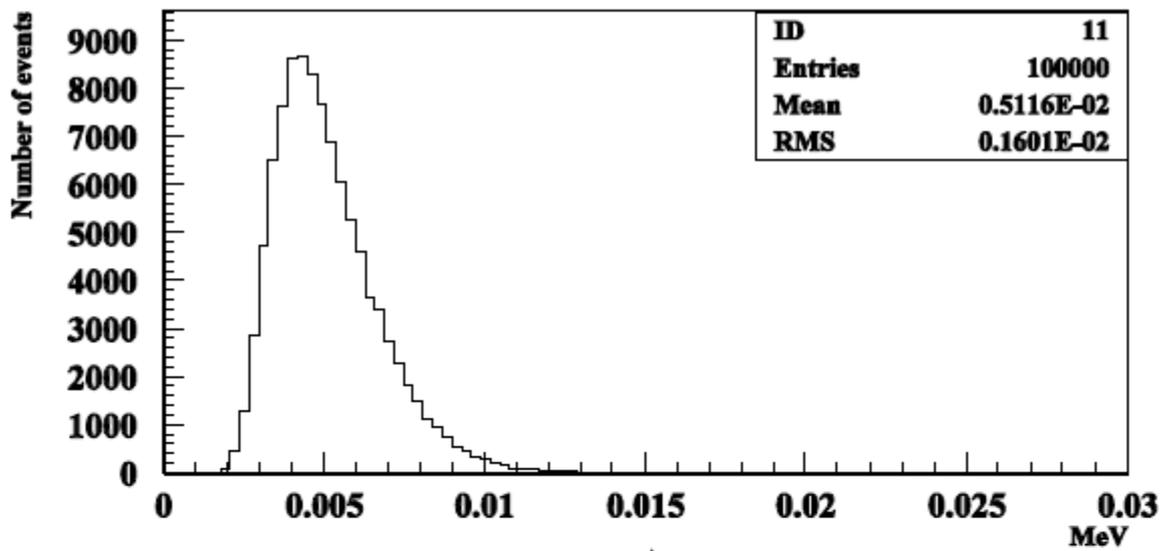

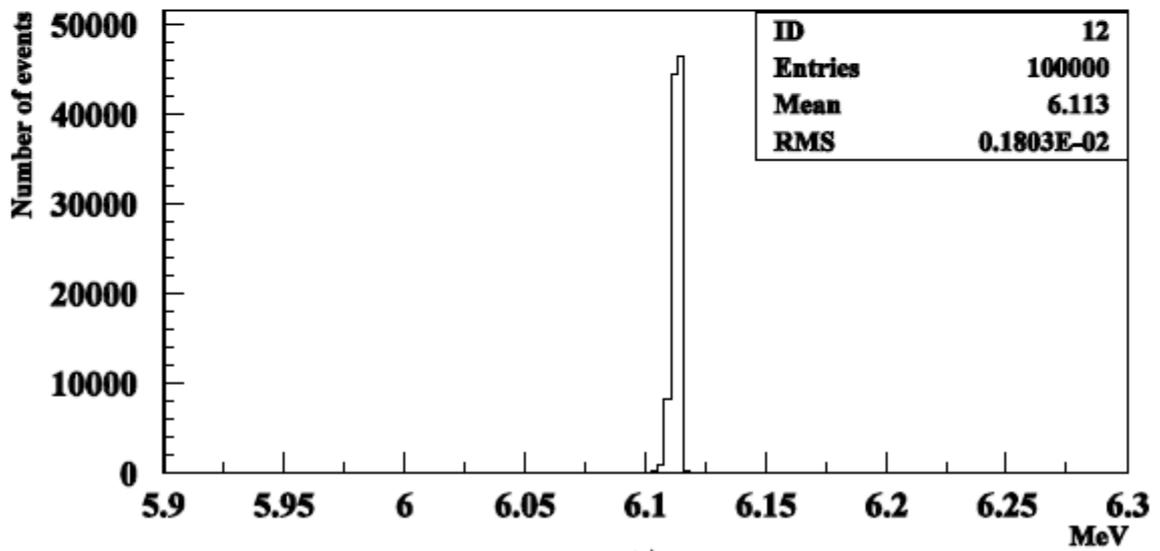

Fig. 6. dE/dx (a) and energy (b) distributions of α-particles from Cf-252 passing through 7.5 µg/cm² carbon.

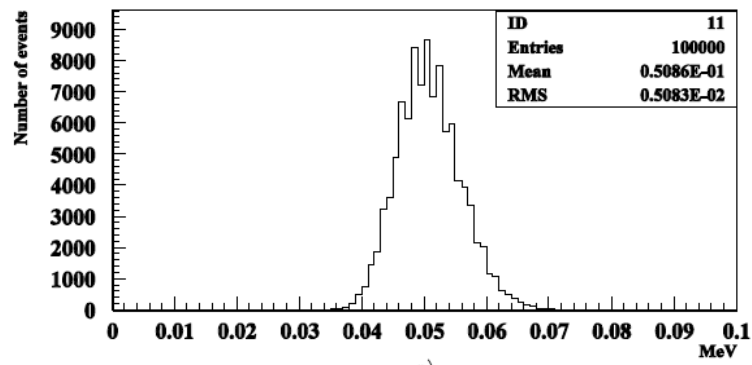

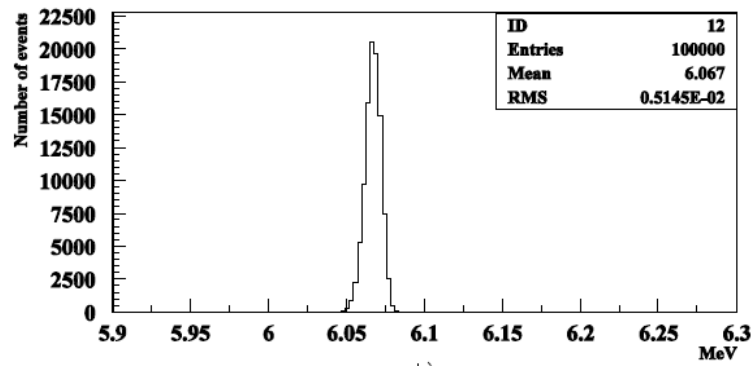

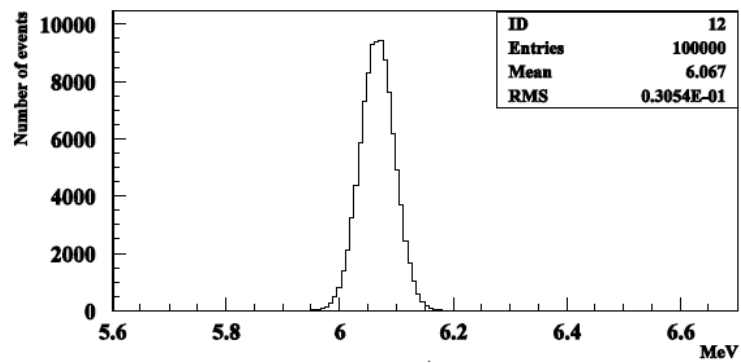

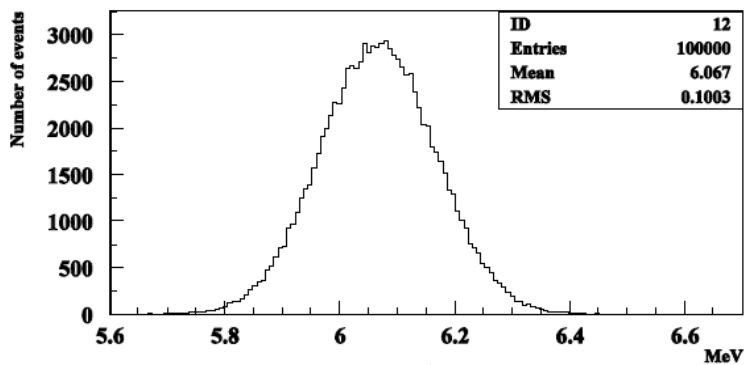

Fig. 7. dE/dx (a) and energy (b) distributions of α-particles from Cf-252 after passing through 75 μg/cm$^2$ carbon. The energy distributions of α-particles detected with 30 and 100 keV resolutions are presented in (c) and (d) respectively.

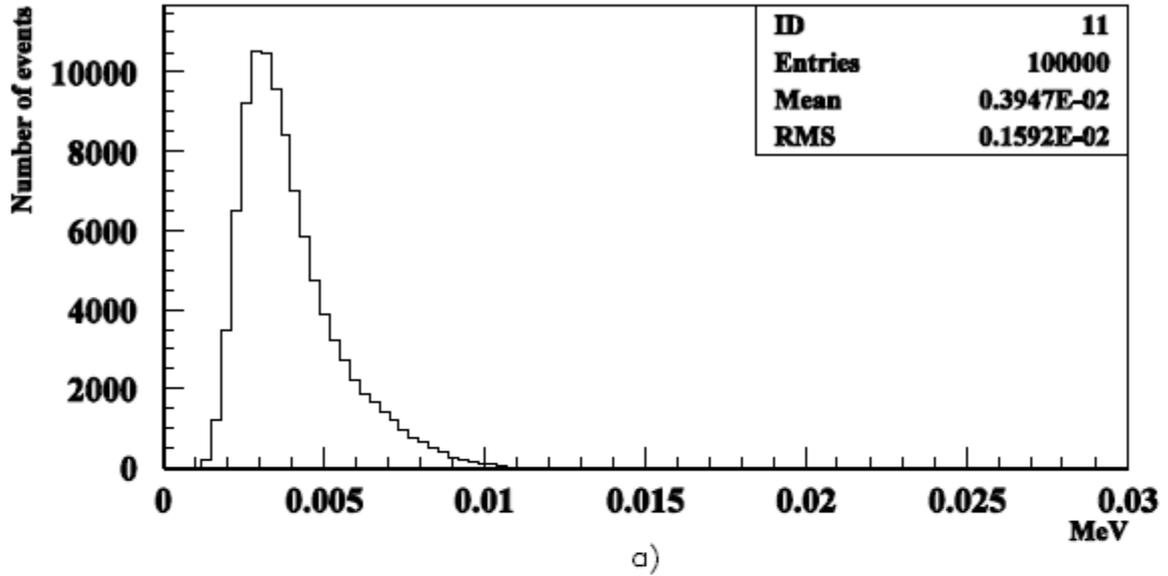

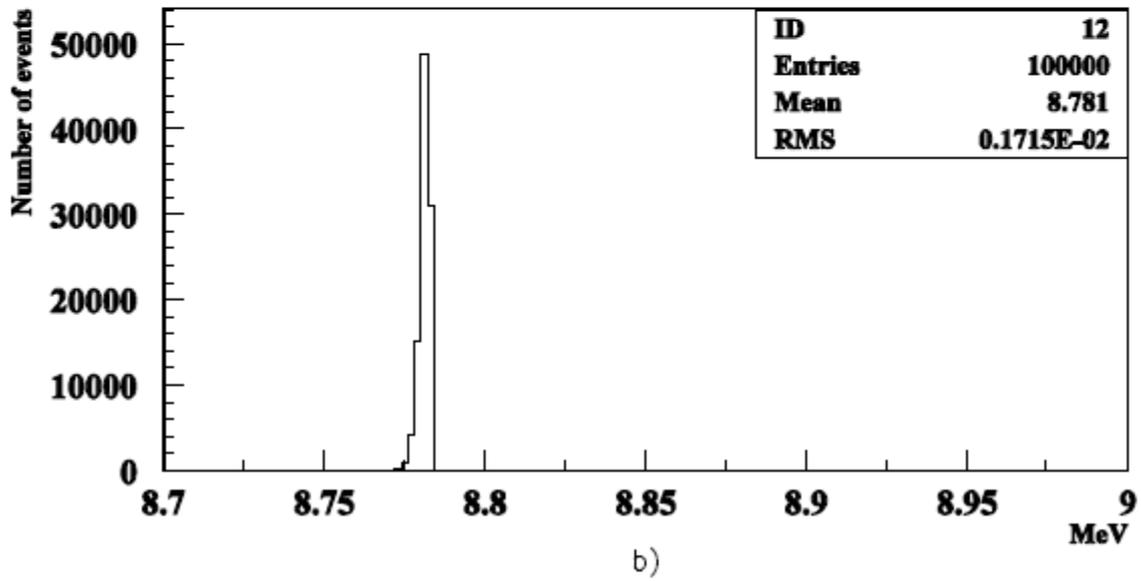

Fig. 8. dE/dx (a) and energy (b) distributions of 8.785 MeV α-particles passing through 7.5 µg/cm² carbon.

The same distributions, but for 8.785 MeV α-particles, are presented in Fig. 8. It follows from these distributions, that if one uses a 2.0-10.0 keV deposited energy equvalent signals from the low-pressure MWPC as a trigger, then practically all α-particles from $^{252}$Cf decays could be tagged. Meanwhile the dE/dx distributions of α-particles passing 75 µg/cm² carbons, i.e. passing about 10 cm, 2 Torr hexane, have a Gaussian shape with a r.m.s. of about 5 keV (see e.g. Fig. 7a). The energy distributions of these alpha particles are mainly determined by the detector resolution, as demonstrated in Fig. 7 b,c,d. Therefore, the device depicted at Fig. 3 can be used to separate α-particles from FFs, β, γ and cosmic rays and carry out precise and sensitive α-particle spectroscopy of $^{252}$Cf decays. In Fig. 9 the expected energy distributions of α-particles from the main decay channel of $^{252}$Cf (15.7% for 6.076 MeV and 84.3% for 6.118 MeV) and 7.2 MeV α-particles from the $^{244}$Cf decay, which is formed after octoneutron emission from $^{252}$Cf, are shown. The octoneutron or 7.2 MeV α-particle emission probability is taken equal to P($^8$n)/P(α) = 1.74 · 10$^{-6}$. The total number of generated events is equal to 10$^7$, from which in 4407 cases energies of α-particles are higher than 6.4 MeV but less than 6.95 MeV, while 19 events around 7.2 MeV are related to the octoneutron emission. These events are presented separately in Fig. 9b.

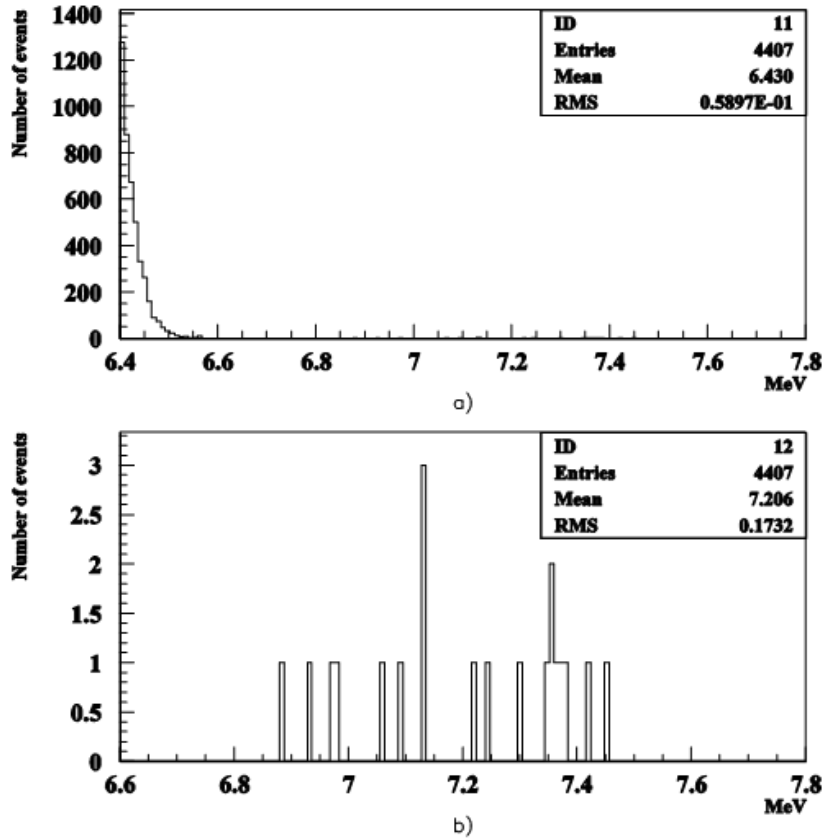

Fig. 9. The energy distributions of α-particles from main decay channel of $^{252}$Cf (15.7% for 6.076 MeV and 84.3% for 6.118MeV) and 7.2 MeV α-particles from $^{244}$Cf, which are formed after octoneutron emission, passing through 75 μg/cm$^2$ carbon. The α-particles are detected with 100 keV resolution. Total number of events is 10$^7$. The number of α-particles with energies larger than 6.4 MeV is equal to 4407, out of which 19 events are due to the octoneutron emission.

As it was mentioned above, another inevitable signature of the octoneutron radioactivity of $^{252}$Cf is the emission of $E_\alpha$ = 6.778 and 8.785 MeV α-particles from the decay of daughter nuclei $^{216}$Po and $^{212}$Po respectively. The relative intensity of these lines is rather sensitive to the age of the $^{252}$Cf source. For instance, for a 25 year old source (to be used in the proposed experiment) their expected intensities exceed that for the $E_\alpha$ = 7.2 MeV line by factors of 20.8 and 13.4, respectively. Per 10$^7$ usual α-decays of $^{252}$Cf, the expected numbers of high-energy α-particles are 365 and 234 for the $E_\alpha$ = 6.778 and 8.785 MeV lines, respectively, compared to about 18 α-particles for the $E_\alpha$ = 7.2 MeV line. The high energy part of the predicted spectrum including 6.29, 6.778, 7.2 and 8.785 MeV groups of α-particles is depicted in Figure 10. It should be noted, that the contribution to this spectrum from the decay chain depicted in Fig. 2 is negligible (less than 1%) due to the very large decay time of $^{232}$Th (T½ = 1.4·10$^{10}$ years) which leads to a strong suppression of the following decay processes.

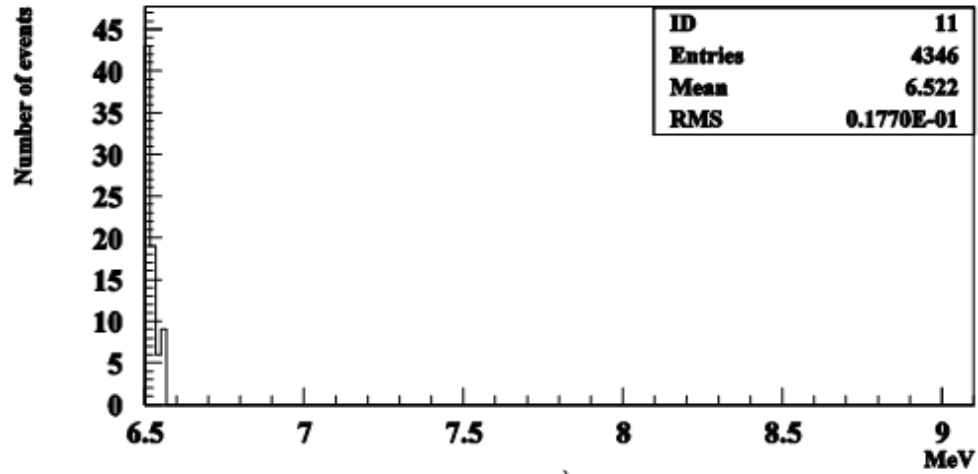

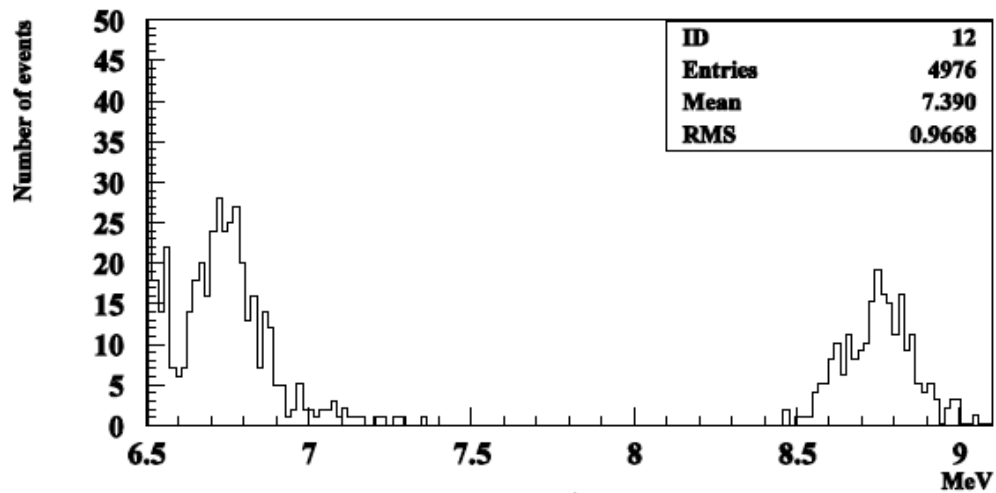

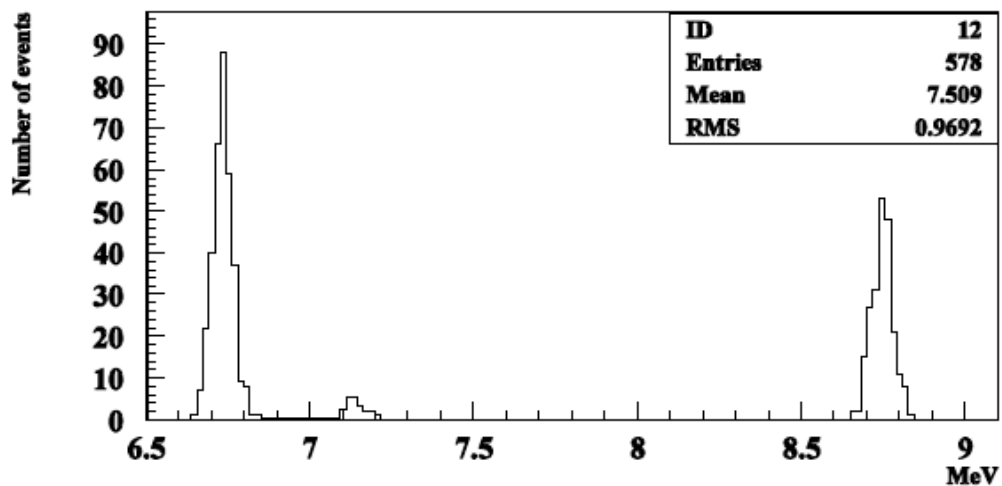

Fig. 10. The high energy part of the energy distribution of α-particles: (a) from the main decay channel of $^{252}$Cf (15.7% for 6.076 MeV and 84.3% for 6.118 MeV); (b and c) 6.29, 6.778, 7.2 and 8.785 MeV α-particles from the decay chain of $^{244}$Cf, which is formed after octoneutron emission, are added. The energy resolution is taken to be 100 keV for figures (a) and (b) and 30 keV for figure (c). Total number of events of α-particle decays is equal to $10^7$.

## 5. Summary

We propose to carry out the α-particle spectroscopy of $^{252}$Cf nucleus by using the experimental setup depicted in Fig. 3. If, indeed, $^{252}$Cf nuclei emit octoneutrons with relative probability $P(^8n)/P(\alpha) = 1.74 \cdot 10^{-6}$, we will be able, using available $^{252}$Cf source at Yerevan Physics Institute, to detect in a six week period about 1000 α-particles from the decays of daughter nuclei $^{244}$Cf, $^{216}$Po and $^{212}$Po originating from the octoneutron radioactivity of $^{252}$Cf. The proposed experimental setup can be also used for detection and identification of other channels of the $^{252}$Cf disintegration, in particular, for direct observation of the cluster decay channels (2)-(10).

**Acknowledgment.** The activity of one of the authors (H.G.) was supported by State Committee Science MES RA, in frame of project № 13-1C245.

**Refeernces**

1. V. A. Varlachev, A. A. Garapatskii, G. N. Dudkin et al., Search for heavy neutron clusters in nuclear fission, Bulletin of the Russian Academy of Sciences: Physics 73 (2), 143 (2009).
2. B. G. Novatsky, S. B. Sakuta, D. N. Stepanov, Detection of light neutron nuclei in the alpha-particle-induced fission of $^{238}$U by the activation method with $^{27}$Al, JETP Letters 98 (11), 656 (2013).
3. G. N. Dudkin, A. A. Garapatskii, V. N. Padalko, Method of searching for neutron clusters, Nucl. Instr. Methods. A 760, 73 (2014).
4. A. Margaryan, J.-O. Adler, J. Brudvik et al., Low-pressure MWPC system for the detection of alpha-particles and fission fragments, Armenian Journal of Physics, 3 (4), 282 (2010).
5. K. A. Ispirian, A. T. Margarian, A. M. Zverev, A Monte-Carlo method for calculation of the distribution of ionization losses, Nucl. Instr. Methods. V. 117, 125 (1974).